\documentclass[%
 preprint,
 tighten lines,
 amsmath,amssymb,
 aps,
prb,
]{revtex4-1}
\usepackage{graphicx}
\usepackage{dcolumn}
\usepackage{bm}
\def\degree{${}^{\circ}$}
\usepackage{epsfig}

\begin{document}


\title{Characterizing thermal conduction in polycrystalline graphene}

\author{Yanlei Wang}
\author{Zhigong Song}
\author{Zhiping Xu}%
 \email{xuzp@tsinghua.edu.cn}
\affiliation{
Applied Mechanics Laboratory, Department of Engineering Mechanics, and Center for Nano and Micro Mechanics, Tsinghua University, Beijing 100084, China
}%

\date{\today}

\begin{abstract}
Thermal conduction was explored and discussed through a combined theoretical and simulation approach in this work. The thermal conductivity $\kappa$ of polycrystalline graphene was calculated by molecular dynamics simulations based on a hexagonal patch model in close consistence with microstructural characterization in experiments. The effects of grain size, alignment, and temperature were identified with discussion on the microscopic phonon scattering mechanisms. The effective thermal conductivity is found to increase with the grain size and decrease with the mismatch angle and dislocation density at the grain boundaries. The $\sim T^{-1}$ temperature dependence of $\kappa$ is significantly weakened in the polycrystals. The effect of grain boundaries in modifying thermal transport properties of graphene was characterized by their effective width and thermal conductivity as an individual phase, which was later included in a predictive effective medium model that showed degraded reduction in thermal conductivity for grain larger than a few microns.
\begin{description}
\item[PACS numbers]
81.05.ue, 65.80.Ck, 61.72.Lk, 61.72.Mm, 44.10.+i
\end{description}
\end{abstract}
\maketitle

\section{\label{sec:level1}Introduction}
Graphene, according to its unique thermal and mechanical properties,\cite{balandin2008superiore1} becomes a promising material for applications such as heat spreading material in very large-scale integration (VLSI),\cite{grosse2011pop} thermal interface materials\cite{wang2013thermal,shahil2012graphene} and so on. In experiments, the room-temperature thermal conductivity $\kappa$ was measured up to 5300 W/mK for a monolayer graphene by inspecting the dependence of the frequency of Raman G peak on the excitation laser power.\cite{balandin2008superiore1} This value is much higher than that of the best bulk thermal conductor, diamond (with $\kappa$ in the range of 1000-2200 W/mK at room temperature),\cite{shinde2006high} and even higher than carbon nanotubes with $\kappa$ = 3500 W/mK at room temperature.\cite{pop2006thermal_cnt} However, large-scale single-crystalline graphene is very difficult to be produced.\cite{yakobson2011observational} Although 30-inch continous graphene films can be grown by the chemical vapor deposition (CVD) technique,\cite{bae2010naturenano} the polycrystalline nature\cite{huang2011grains_cvd} of these films poses a question that how the thermal transport properties of graphene is modified by the presence of lattice imperfections, especially the grain boundaries (GBs).

The thermal conductivity of graphene at room temperature was predicted theoretically with values ranging from a few hundreds to 8000 W/mK, depending on the interatomic potentials and simulation methods adopted.\cite{zhang2010GK1,wei2011apply,evans2010thermal} On the other hand, experimental measurements suggested values from $4.84\times {10}^{3}$ to $5.30\times {10}^{3}$  W/mK, where the material actually features a polycrystalline nature with grain size of a few to hundreds of microns.\cite{balandin2008superiore1,huang2011grains_cvd,geng2012hex1} It is not clear yet how the polygrain nature could modify the intrinsic thermal transport properties of graphene. In the literature, the GB was known to play a key role in determining $\kappa$ for semiconductors. Superlattice structures could reduce the thermal conductivity of material to values below its alloy limit, in the mechanism of minimum thermal conductivity where the combined role of the particle and wave nature of phonon arises.\cite{garg2013minimum} For example, $\kappa$ of the GaAs/AlAs superlattice was reported to be reduced by about one order of magnitude compared to the bulk conductivity of GaAs,\cite{capinski1996tc_superlattice} this significant reduction was observed experimentally and also analyzed by several models.\cite{chen1998superlattice2,superlattice1} With these evidences in mind, it is thus interesting and of practical importance to assess the effect of GBs on thermal conduction in graphene.

There were several studies exploring thermal transport properties of GBs in graphene. For example, Bagri \emph{et al.}\cite{sheony2011thermalth2} studied the thermal transport across twin GBs in polycrystalline graphene using molecular dynamics (MD) simulations and found that the boundary conductance is significantly higher than that of other thermoeletric interfaces. Cao \emph{et al.}\cite{cao2012kapitza} have studied the temperature, grain size, and alignment dependence of $\kappa$ for graphene with parallel GBs. A more recent study explored the effect of grain size, or dislocation density in the GB, more quantitatively, and showed that $\kappa$ is not significantly reduced at room temperature if the grain size reaches several microns.\cite{serov2013effect} However, all these studies were focusing on graphene with parallel GBs, which obviously deviate from the structure of polycrystalline graphene identified in experiments.\cite{huang2011grains_cvd} In this work, we will study $\kappa$ of polycrystalline graphene consisting of hexagonal grains, as a more realistic model, and the results are expected to be feasible to compare with experimental measurements.\cite{kotakoski2012model1,van2013model2,song2013pseudo_model3,geng2012hex1,wu2012hex2}

MD simulations are usually limited by the system size it could model. To extend our discussion to the material level, an effective medium theory (EMT) where the polycrystalline graphene is considered as a two-phase composite was also used to predict thermal transport properties up to the micrometer scale.\cite{nan1997_composite,hasselman1987_composite2} This approach was widely used to identify thermal conductivities of material interfaces and nanocomposites. For example, Nan and his collaborators introduced a so-called Kapitza radius for the interfacial thermal resistance via applying the Maxwell-Garnett effective medium theory to predict thermal properties of nanocomposites.\cite{nan1997_composite} Hao \emph{et al.}\cite{hao2012thermal} studied the effective thermal conductivity of crystalline Si-Ge nanocomposites using EMT by considering the interface between Si and Ge as an individual phase. We followed this approach in this work to analyze our results.

The paper is arranged as follows. After this introduction, which presents the background and motivations, section II introduces the atomic structure of polycrystalline graphene and the methodological details we used in MD simulations. Then in section III, the dependence of $\kappa$ for polycrystalline graphene on the grain size, alignment and temperature are described and analyzed. In section IV, the determination of $\kappa$ for polycrystalline and the major factors affecting $\kappa$ are discussed, as well as the temperature dependence, through the microscopic mechanisms of thermal conduction. Finally, main conclusions are summarized in section V.
\section{\label{sec:level1}Models and methods}
\subsection{\label{sec:level2}The atomic structures of polycrystalline graphene}
Recently, films of uniform hexagonal graphene flakes were synthesized on liquid copper surfaces, which held great promises in growing large-scale graphene films with uniform and well-controlled grain size and shape.\cite{geng2012hex1,wu2012hex2} In this work, we created model structures of polycrystalline graphene from hexagonal grains following this concept, instead of randomly patching graphene flakes with irregular shapes into a 2D membrane.\cite{kotakoski2012model1,huang2011grains_cvd} These graphene hexagonal flakes were defined by its grain size \emph{L} and orientation angle $\theta$, as denoted in Fig. 1. In our work, the grain size \emph{L} is defined as the edge length of a hexagonal grain, and the orientation angle $\theta$ was defined as the angle between zigzag direction and $X$-axis in our coordinate system. Considering the hexagonal symmetry of the graphene lattice, a modulated value of $\theta$ between 0\degree\ and 30\degree\ was used, i.e. $\theta$ and 60\degree-$\theta$ are equivalent. Thus a whole polycrystalline model can be denoted by a triple array of angles ($\alpha$, $\beta$ and $\gamma$), that defines the orientation alignment between neighboring grains. To describe the orientation of the whole polycrystalline material, we defined a relative orientation angle (ROA) ${\theta}_{r} = (|\alpha - \beta|+|\alpha - \gamma|+|\beta - \gamma|)/3$. We can denote the misorientation angle of the two crystalline dominions for a GB in graphene\cite{yazyev2010topological} and here the ROA is the mean value of the three orientational angles in polycrystalline graphene, so we can use ROA to describe the grain alignment of polycrystalline graphene. Successive procedures to optimize the GBs were carried out further by adding, removing carbon atoms, and adjusting local bonding topologies, to ensure the relative energy of each carbon atom to the value in bulk graphene less than 1 eV, corresponding to an energetically well-defined dislocation array in the GB.

In principle, this approach could be used to create polycrystalline graphene with any grain size, shape and ROA. But here, without loss of the generality, we focused on six polycrystalline graphene structures with hexagonal grains. Their angle triplet ($\alpha$, $\beta$ and  $\gamma$) are (0\degree, 5\degree, 10\degree), (0\degree, 5\degree, 15\degree), (0\degree, 5\degree, 20\degree), (0\degree, 10\degree, 20\degree), (0\degree, 10\degree, 25\degree), (0\degree, 15\degree, 30\degree) as denoted in Fig. 1(a), and the corresponding ROAs are 5\degree, 10\degree, 13.33\degree, 13.33\degree, 16.67\degree\ and 20\degree, respectively. Although not investigated in this work, other combinations of grain orientation angles are expected not to change the physics under discussion here qualitatively. The grain size effect for polycrystalline graphene was characterized by considering models with \emph{L} = 1, 2, 3 and 5 nm. It should be noted that graphene with grain size smaller than 1 nm we investigated is not discussed in this paper as the high density of defects therein lead to diminishing difference between the grain and GB, and thus ambiguous definitions of their sizes.
\subsection{\label{sec:level2}Computational methods}
To compute the thermal conductivity $\kappa$ from MD simulations, we used the Green-Kubo method based on the linear response theory,\cite{zhang2010GK1,hao2011GK4} which applies for systems in thermal equilibrium where heat flux fluctuates around zero. $\kappa$ could thus be expressed as an integration of the heat flux operator multiplied by a prefactor
\begin{equation}
{\kappa }_{xy}=\frac{1}{V{k}_{\mathrm{B}}{T}^{2}}\int_{0}^{{\tau}_\mathrm{c}}<{J}_{x}(\tau)\cdot {J}_{y}(0)>\mathrm{d}\tau
\end{equation}
where $T$ is the temperature of system, ${k}_{\mathrm{B}}$ is the Boltzmann constant, and $V$ is the system volume that was defined here as the area of polycrystalline graphene multiplied by its nominal thickness (the inter layer distance of graphite, 3.4 $\mathrm{\AA}$). The upper limit of time integration ${\tau}_{c}$ needs to be long enough so the current-current correlation decays to zero,\cite{schelling2002GK2} ${J}_{x}$ and ${J}_{y}$ are the heat current operators in the $x$ and $y$ directions, and the angular bracket represents the ensemble average,\cite{hardy1963heatflux} namely the heat flux autocorrelation function (HFACF). The heat flux \textbf{J} of the system was computed from the expression $\textbf{J}=(\sum_{i} {e}_{i} {\textbf{v}}_{i} - \sum_{i} {\textbf{S}}_{i} {\textbf{v}}_{i})/V$, where ${e}_{i}$, ${\textbf{v}}_{i}$ and ${\textbf{S}}_{i}$ are the total energy, velocity vector, and stress tensor of each atom $i$, respectively. We first integrated HFACF with an integration time $\tau$ following Eq. (1), then obtained the relation between $\kappa$ and $\tau$. The decorrelation time ${\tau}_{c}$ for the heat flux is typically on the scale 10 ps for our models, and thus converged results for $\kappa$ could be extracted when $\tau$ $>$ ${\tau}_{c}$ in the simulations. The advantage of Green-Kubo method, compared to the non-equilibrium molecular dynamics (NEMD) simulations with thermal gradient built up across the material, is that it does not require additional perturbation to the equilibrium that may give rise to nonlinearity, and has weaker dependence on the size of simulated system. Due to the hexagonal symmetry of graphene lattice, the thermal conductivity of polycrystalline graphene is isotropic. We thus evaluate $\kappa$ as the mean value of ${\kappa}_{xx}$ and ${\kappa}_{yy}$ that may differ in the finite system under simulation though, where the maximum difference between ${\kappa}_{xx}$ and ${\kappa}_{yy}$ in our simulations is less than $0.1\kappa$.

All simulations were performed using the large-scale atomic/molecular massively parallel simulator (LAMMPS).\cite{plimpton1995lammps} Periodic boundary conditions were applied to a 2D supercell of polycrystalline graphene, and in order to minimize the size effect, we used almost the same supercell size in the MD simulations (18.0$\times$312.0 ${\mathrm{nm}}^{2}$ for polycrystalline graphene with $L=$ 1, 2, 3 nm, 15.0$\times$260.0 ${\mathrm{nm}}^{2}$ for polycrystalline graphene with $L=$ 5 nm, and 17.9$\times$312.0 ${\mathrm{nm}}^{2}$ for pristine graphene). The optimized Tersoff potential,\cite{lindsay2010optiTersoff} which has been demonstrated to provide an excellent prediction for the phonon dispersion graphene in comparison with experimental measurements, was used for the interatomic interactions between carbon atoms. This method was widely used to predict thermal and mechanical properties of graphene and related materials.

In the MD simulations, the atomic structures of polycrystalline graphene were firstly equilibrated at ambient condition (temperature $T$ = 300 K, where the quantum correction is negligible\cite{savin2012semiquantum}) under a Nos\'{e}-Hoover thermostat for 200 ps, in which the time step is 0.2 fs. Then the structure was further equilibrated in a microcanonical ensemble for 50 ps before the thermal conductivity was calculated. In the equilibrium Green-Kubo calculations, the system was simulated in a microcanonical ensemble as well. The atomic positions and velocities were collected along the simulations to evaluate the heat flux and its autocorrelation functions. The thermal conductivity was finally obtained by following the Green-Kubo formula and averaging over eight independently sampled runs for each structure.
\section{\label{sec:level1}Results}
\subsection{\label{sec:level2}The dependence of $\kappa$ on grain size and orientation}

We first explored the grain size dependence at \emph{T} = 300 K. Data points plotted in Fig. 2(a) are our calculation results for $\kappa$ of polycrystalline graphene with grain size ranging from 1 to 5 nm, but the same orientation (0\degree, 5\degree, 10\degree). The results show that $\kappa$ increases for larger size of grains. The value for $\kappa$ is 60.76 W/mK for $L$ = 1 nm and 197.69 W/mK for $L$ = 5 nm. The data can be well fitted into a linear function as ${\kappa}=26.67+35.24L\ (\mathrm{nm})\  \mathrm{W/mK}$. For comparison, the thermal conductivity of graphene was calculated to be 892.68 W/mK following the same simulation procedure. Thus the incorporation of nanosized grains reduces the thermal conductivity of graphene remarkably, by amount up to 93 percent (for $L$ = 1 nm). Further calculations for models with different ROAs yield similar results.

In experiments, GBs with different orientational mismatch were identified,\cite{huang2011grains_cvd} that corresponds to a certain density level of dislocations or defects in the GB. Considering the GB as a linear array of edge dislocations, the density of consecutive dislocations is thus given by the Frank formula, $\rho$ = 1/$d$ = 2$\mathrm{sin}$($\theta$/2)/$|\textbf{b}|$, where \textbf{b} is the Burgers vector and $d$ is the distance between two neighboring dislocations.\cite{song2013pseudo_model3} To quantify the effect of orientation angles, we calculate values of $\kappa$ for polycrystalline graphene with different ROAs. The results are shown in Fig. 2(b), which suggest that $\kappa$ decreases as the ROA ${\theta}_{r}$ increases, and the trend is the same for different grain size. For $L$ = 5 nm, $\kappa$ is 197.69 W/mK when the angle triplet is (0\degree, 5\degree, 10\degree) and 126.09 W/mK for the angle triplet of (0\degree, 15\degree, 30\degree), which is reduced by 36 percent as ${\theta}_{r}$ increases from 5\degree\ to 20\degree.

\subsection{\label{sec:level2}Heat flux distribution and the effective width of the grain boundary}
The observation that effective thermal conductivity of the polycrystalline graphene $\kappa$ increases with the grain size $L$ and decreases with the ROA ${\theta}_{r}$ is closely related to the atomic structure of polycrystalline graphene. The structure of a GB in our model comprises of a series of topological defects, e.g. pentagon-heptagon (5$|$7) pairs, and the GB tends to break into an array of dislocation as shown in Fig. 1(b). The properties of polycrystalline graphene thus will rely on the characteristics of the GB (size, the density of defects, etc.). The role of dislocation in modifying mechanical and thermal properties was well identified for bulk materials,\cite{ziman2001phonon,hull2001introduction} as material discontinuity in crystals. From the continuum theory where the dislocation is seen as a line in 3D or a point in 2D, one can define the effective width of the dislocation as\cite{ziman2001phonon}
\begin{equation}
\emph{w} = \frac{1}{2}{\gamma}^{2}{b}^{2}q \sim \pi{\gamma}^{2}{b}^{2}/\lambda
\end{equation}
where the length of the Burgers vector $b$ is usually about one lattice constant, \textbf{q} and $\lambda$ are the wave vector and wavelength of the phonon, and the parameter $\gamma$ is a dimensionless parameter that describes the relation between dilatation and temperature when the temperature is low. Considering the dislocation core, i.e. the region where the atomic displacements are so large that elasticity theory can not adequately applied, one can follow the theory of Rayleigh scattering and estimate the scattering width as\cite{ziman2001phonon}
\begin{equation}
\emph{w} = a{(\frac{\Delta D}{D})}^{2}{(qa)}^{3}
\end{equation}
where $a$ is the radius of the scattering region where the material has a constantly shifted density (from $D$ to $D+\mathrm{\Delta} D$). Although the assumptions in Eqs. (2) and (3) are very crude, it reveals some basic aspects such as that the effective width of dislocation is finite and is quite small for long-wave phonons.

If a crystal contains a GB of dislocations, a propagating phonon attempting to cross the GB may find itself in a region where it can no longer proceed in the same direction with the same velocity and energy. That is to say, just as a single dislocation has an finite scattering width, there must also be a bounded region with an effective width \emph{w}. Inside this region the crystal structure can not be treated as a perfect one. To find more details about this bounded region as well as the definition of $w$, and to obtain more insights into the grain size and orientation dependence of thermal conductivity for polycrystalline graphene, we plot the spatial distribution of heat flux within the whole polycrystal in Fig. 3, which displays the heat flux of polycrystalline graphene for the specified angle triplet (0\degree, 5\degree, 10\degree). This was done by calculating atomic heat flux in NEMD simulations that was averaged over 10 ps in the steady state. The atomic heat flux is defined from the expression ${\textbf{J}}_{i}={e}_{i} {\textbf{v}}_{i}-{\textbf{S}}_{i} {\textbf{v}}_{i}$. The results clearly identified strong scattering of the heat flux in the region adjacent to the GBs. However, inside the grains and away from the GBs, the spatial distribution becomes close to that in a single crystal of graphene. When the grain size increases, the portion of scattered region becomes reduced. Thus we can consider the polycrystalline graphene as a composite consisting of two phases - the pristine grains and GBs that feature quite different material properties. Furthermore, from the heat flux distribution, we could conclude visually that the scattering region spans in few lattice constants, and quantitative determination of $w$ will be introduced below with respect to a predictable model for thermal conductivity of polycrystalline graphene.

\subsection{\label{sec:level2}Effective medium theory for the prediction of $\kappa$}
A validated effective model could be used to predict thermal transport properties of polycrystalline graphene with grain size much larger than the one our MD simulations are capable to capture. The key to establish such a model is to well define the model and fit parameters therein. In our polycrystaline graphene models, there are three GBs with different misorientation angles for each sample, and thus the effective width $w$ and thermal conductivity $\kappa_\mathrm{GB}$ for the GB phase should be different with respect to these GBs. However, we noted in the aforementioned simulation results (Fig. 2(b)) and previous reports\cite{cao2012kapitza} that there is only small difference in $\kappa_\mathrm{GB}$, e.g. $\kappa_\mathrm{GB}$ for a zigzag GB with orientation angle 21.7\degree\ is only 15 percent less than that of a zigzag GB with orientation angle 5.5\degree. Moreover, by assumption these three GBs have the same volume fraction and distribute homogeneous in our EMT model although the density of dislocations is different, which can be seen from Fig. 1. Thus for simplicity, we only divided the polycrystalline graphene into two different phases, i.e. the bulk phase with the properties as  pristine graphene, and a GB phase with no microstructures specified. In other words, we considered the material as a 2D composite with GBs embeded in bulk graphene as inclusions. The overall thermal resistance of the composite, $\kappa$ could thus be predicted as\cite{hao2011GK4}
\begin{equation}
 \frac{1}{\kappa} = \frac{f_{\mathrm{GB}}}{{\kappa}_{\mathrm{GB}}} + \frac{1-f_{\mathrm{GB}}}{{\kappa}_{\mathrm{G}}}
\end{equation}
where $\kappa$, ${\kappa}_{\mathrm{GB}}$ and ${\kappa}_{\mathrm{G}}$ are the predicted effective thermal conductivities of the composite, the GB and the bulk phase (pristine graphene), and $f_{\mathrm{GB}}$ is the volume fraction of the GB phase. In order to use Eq. (4) for quantitative prediction, we need firstly to determine the effective width of the GB so $f_\mathrm{GB}$ could be defined, through $f_\mathrm{GB} = (2\sqrt3wL-{w}^{2})/{3{L}^{2}}$.

But how could we determine \emph{w} from the MD simulation results? As mentioned previously, we defined $w \approx 0.7$ nm visually by inspecting the spatial distribution of heat flux and measuring the size of scattering region in the Fig. 3. On the other hand, the value thus determined could be verified by substituting the value and $f_\mathrm{GB}$ it defines into Eq. (4) to see if the simulation results for $\kappa$ could be well fitted. In statistics, the coefficient of determination,\cite{carpenter1960principles} denoted as ${R}^{2}$, indicates how well data points fit a line or curve. In our fitting, ${R}^{2}$ is up to 0.99, so we can say that Eq. (4) fits the simulation results very well. From now on, without loss of accuracy, we can define the effective width \emph{w} constantly as 0.7 nm in our EMT model. Fitting the simulation results in Fig. 4(a) gives:
\begin{equation}
{\kappa} = \frac{719.97}{1+16.81f_{\mathrm{GB}}}\   (\mathrm{W/mK})
\end{equation}
Based on Eq. (5), we fitted the parameters in the EMT model, we can then predict $\kappa$ for polycrystalline graphene with arbitrary grain size as the physics is not changed. The result shows that as $L$ increases, $\kappa$ increases and converges to a limit ${\kappa}$ = 719.97 W/mK that corresponds to the value for pristine graphene. The $L$-dependence of $\kappa$ for polycrystalline graphene with the specified angle triplet (0\degree, 5\degree, 10\degree) is plotted in Fig. 4(b), as well as Serov et al.'s calculation results for polycrystalline graphene with parallel GBs.\cite{pop2013effectth1} The values were both normalized by the thermal conductivity of pristine graphene, i.e. the limit at infinitely large grain size. The comparison shows that although the trend is in general the same, the predicted effective thermal conductivity for our polycrystalline model with hexagonal grains is higher than the superlattice-like model explored by Serov.\cite{pop2013effectth1} The difference may thus be attributed to the contrastive structure (the density of dislocations, orientation angle, etc.) and scattering of phonons propagating across the GBs perpendicularly in their work. Our results for the more realistic polycrystalline models suggest that at room temperature, the thermal conductivity of polycrystalline graphene with grain size larger than several hundred of nanometers shows no significant reduction in comparison with the pristine single crystal.

\subsection{\label{sec:level2}The temperature dependence of $\kappa$}
The temperature dependence of $\kappa$ for graphene and polycrystalline graphene calculated in our studies are shown in Fig. 5. For comparison, we also plot $\kappa$ measured for suspended CVD-grown graphene at different temperature (by Cai \emph{et al.}\cite{cai2010thermale2}) and calculated for polycrystals with parallel GBs (by Cao \emph{et al.}\cite{cao2012kapitza}) in Fig. 5. For all the results, we found that $\kappa$ monotonically decreases with the temperature that agrees well with other theoretical studies, for example Ref. 46. The results in Fig. 5 suggest a temperature dependence of $\kappa \sim {T}^{-\alpha}$ for $T$ from 200 K to 1000 K for all the cases, and we can define an exponent \emph{$\alpha$} to capture this scaling behavior. $\kappa$ for pristine graphene calculated in this work shows a scaling behavior of $ \sim {T}^{-0.96}$, while $\kappa$ for experimental measurements is $ \sim {T}^{-1.03}$. There is thus $ \sim$ 7 percent difference in the exponents \emph{$\alpha$} determined from our simulations compared to experimental measurements. The value of \emph{$\alpha$} is reduced when the polycrystalline nature is introduced. For the graphene with parallel GBs,\cite{cao2012kapitza} \emph{$\alpha$} $ \sim$ 0.79, and for our polycrystalline graphene models with hexagonal grains, the exponents \emph{$\alpha$} vary from 0.22 to 0.32 with the grain size increases from 1 to 5 nm, which are both much smaller in comparison with \emph{$\alpha$} $ \sim$ 0.96 for the pristine graphene. From the scaling relation $\kappa \sim {T}^{-\alpha}$, we conclude that the temperature dependence of $\kappa$ is significantly reduced for polycrystalline graphene, which is evidenced in both the reduced exponent \emph{$\alpha$} compared to the pristine graphene and enhanced \emph{$\alpha$} at large grain size.

\section{\label{sec:level1}Discussion}
\subsection{\label{sec:level2}Phonon scattering at GB and the effective width}
Considering the graphene with an armchair GB with orientation angle 15.4\degree\ as shown in the Fig. 6(a), the propagating phonon will experience reflection and refraction when it transports across the GB and its group velocity will have a mismatch between the two crystalline domains on each side of the GB. This mismatch suggests additional phonon scattering sources in addition to the lattice distortion and disorder at the GB. To identify the scattering effect on the thermal transport quantitatively, we calculate $\kappa$ for the graphene model with this infinite GB (across the periodic boundary) using the Green-Kubo method described above. The results show that ${\kappa}_{\perp} = 122.03\ (\mathrm{W/mK})$ and ${\kappa}_{\parallel} = 249.68\ (\mathrm{W/mK})$, where ${\kappa}_{\perp}$ and ${\kappa}_{\parallel}$ are $\kappa$ of graphene with GBs when the direction is perpendicular to GB and when the direction is parallel to GB respectively. To further analyze these results, we divided the polycrystalline graphene with single GB into two phase, i.e. the bulk graphene and GB phases. For thermal transport perpendicular or parallel to the GB, we assumed the effective width of GB $w$ as $\emph{w}_{\perp}$ and $\emph{w}_{\parallel}$, and the thermal conductivity $\kappa$ of GB as ${\kappa}_{\mathrm{GB}\perp}$ and ${\kappa}_{\mathrm{GB}\parallel}$, respectively. By applying Eq. (4) we obtain
\begin{equation}
(\frac{1}{{\kappa}_{\mathrm{GB\perp}}}-\frac{1}{{\kappa}_{\mathrm{G}}}){\emph{w}_{\perp}} = 0.07\  (\mathrm{nm\times mK/W})
\end{equation}
\begin{equation}
(\frac{1}{{\kappa}_{\mathrm{GB\parallel}}}-\frac{1}{{\kappa}_{\mathrm{G}}}){\emph{w}_{\parallel}} = 0.03\  (\mathrm{nm\times mK/W})
\end{equation}
As seen from the above relation, the definition of effective width $w$ is found to be coupled with $\kappa$ of GB, which makes the determination of these two parameters from simulation results nontrivial. To solve this problem, we can define that GB has an intrinsic thermal conductivity ${\kappa}_{\mathrm{GB}}$ and assume it as the mean value of ${\kappa}_{\mathrm{GB}\perp}$ and ${\kappa}_{\mathrm{GB}\parallel}$, i.e. ${\kappa}_{\mathrm{GB}}$ = (${\kappa}_{\mathrm{GB}\perp}$ + ${\kappa}_{\mathrm{GB}\parallel}$)/2. This crude treatment, however, is convenient in estimating effective properties of a realistic polycrystalline graphene with two-dimensional patches of grains. For a single GB, the effective width \emph{w} should be defined uniquely and thus $\emph{w}_{\perp}$ should equal to $\emph{w}_{\parallel}$. From the assumption above and calculations following Eqs. (6) and (7), we have \emph{w} = 0.59 nm, which conveys the thickness of single dislocation. In our previous discussion based on the spatial distribution of heat flux in the polycrystalline graphene, the effective width was found to be 0.7 nm, which is close to the value $w$ = 0.59 nm extracted here by considering the fact that both GBs consist of one array of dislocations. The agreement suggests that the assumption we made is reasonable and the definition of GB as an individual phase with well-defined structural and thermal transport properties is feasible in discussing the effective thermal conductivity of polycrystalline graphene.

To analyse our results via the EMT, we have introduced the effective width \emph{w} to calculate the volume fraction of different phases. Naturally, a problem arises that how to determine the effective width of a GB. In the preceding section, we have obtained an effective width from the heat flux field in MD simulations, but now we discuss the definition of effective width directly from the strain or stress field of graphene with GBs. Ziman\cite{ziman2001phonon} argued that for a GB of orientation angle $\theta$ tends to break into an array of dislocations, spaced by $\emph{d} = b/2\mathrm{sin}(\theta/2) \sim b/\theta$. The effect of such GB can be found in the same way as for a single dislocations. It is known that, a GB, spaced by $\emph{d}$ apart in the $y$ direction, gives rise to a shear strain at the point ($x$, $y$)
\begin{equation}
{\varepsilon}_{xy} = \frac{b}{2\pi(1-\mu)}\mathrm{Re}[{\frac{{\pi}^{2}x}{{d}^{2}}{\mathrm{cosech}}^{2}\frac{\pi (x+iy)}{d}}]
\end{equation}
where $\mu$ is the Poisson's ratio, and $x$, $y$ are measured from the dislocation core, as indicated in Fig. 6(c). At a large distance from the GB, this shear strain can be approximately written as a function of $x$
\begin{equation}
{\varepsilon}_{xy}(x) = \frac{b}{2\pi(1-\mu)}\frac{4{\pi}^{2}x}{{d}^{2}}{e}^{-2\pi x/d}
\end{equation}
which decreases exponentially from the GB and can be neglected for $x \gg d$. So we can measure the spatial distribution of shear strain in experiments or simulations, namely ${\varepsilon}_{xy}$. As described above, we can define a specified value \emph{w}, when $x$ is larger than half of this value \emph{w} the shear strain will be small enough to be neglected. Then we can define the specified value \emph{w} as our effective width of GB. This approach can be used to determine the effective width of a GB without referring to thermal transport measurements. The value of \emph{w} thus determined can be further verified by comparing it with the value obtained from MD simulations that was discussed in the previous sections. Here we define the characteristic strain of graphene as ${\varepsilon}_{i} = (({r}_{1}+{r}_{2}+{r}_{3})/3-{r}_{0})/{r}_{0}$, where ${\varepsilon}_{i}$ is the strain of the atom $i$, ${r}_{1}$, ${r}_{2}$, ${r}_{3}$ is the length of the three bonds with atom $i$ and ${r}_{0} = 1.42\  \mathrm{nm} $ is the carbon-carbon bond length in pristine graphene. For the graphene model with parallel GBs as shown in Fig. 6(a), we calculated the spatial distribution of strain field from our MD simulation results and plotted it in Fig. 6(b). By assuming that the strain can be neglected when it's smaller than $4\times{10}^{-2}$, we measured the effective scope from the distribution of strain and obtain that the value is in the range from 0.6 to 1.0 nm, which agrees to the effective width 0.7 nm and 0.59 nm as extracted in the previous sections.

The definition of effective width $w$ also depends on the density of 5$|$7 dislocations, which is related to the orientation angle via the Frank formula.\cite{song2013pseudo_model3} In the polycrystalline graphene we constructed, there are three GBs featuring different density levels of 5$|$7 dislocations. By neglecting the difference in $\kappa_\mathrm{GB}$\cite{cao2012kapitza}, these three GBs were homogenized in our EMT model. Moreover, from the results of ${\kappa}_{\perp} = 122.03\ (\mathrm{W/mK})$ and ${\kappa}_{\parallel} = 249.68\ (\mathrm{W/mK})$ for the armchair GB with orientation angle 15.4\degree, we know that $w$ should also rely on the angle between the direction of heat flux and GBs, namely $\chi$. In our polycrystalline graphene model, $\chi$ is different for GBs at different positions as the heat flux was applied along the $X$-direction. However, no significant difference was found for the spatial distribution of heat flux (as shown in Fig. 3), so we excluded the effect of $\chi$-depdendence when analyzing $\kappa$.

\subsection{\label{sec:level2}The factors that affect thermal transport in polycrystalline graphene}
To see what factors play key roles in defining the thermal transport properties of the polycrystalline graphene, we extend the discussion with analysis based on the Boltzmann transport equation (BTE), which was widely used to solve thermal transfer problems in materials at the carrier level by neglecting phonon coherence. The general form of BTE is
\begin{equation}
(\frac{\partial}{\partial t}+\textbf{v} \cdot \nabla)f(\textbf{r},\textbf{p},t) = {({\frac{\partial f}{\partial t}})}_{\mathrm{scattering}}
\end{equation}
where \textbf{r} is the position, \textbf{p} is the momentum, $t$ is time, \textbf{v} is the particle velocity, and $f$(\textbf{r},\textbf{p},t) is the single particle distribution function in the phase space at given time. The factors that affect the thermal transport properties can be discussed based on this equation. To solve the BTE (Eq. (10)), one can use the single-mode relaxation time (SMRT) approximation,\cite{srivastava1991physics} which assumes that each phonon mode is characterized by its own relax time that is independent of all other phonons in the system. With this assumption, the scattering term in Eq. (10) can be approximated as
\begin{equation}
{({\frac{\partial f}{\partial t}})}_{\mathrm{scattering}} = \frac{-{f}^{'}}{\tau}
\end{equation}
where ${f}^{'}$ is the fluctuation of the distribution function at equilibrium and $\tau$ is the relaxation time.
For the phonon scattering by 5$|$7 dislocations, the relaxation time is related to the defect density $\rho$, the group velocity $v$ and the effective width of dislocation \emph{w} as\cite{kim2006phonon}
\begin{equation}
\tau \sim \frac{1}{\rho v w}
\end{equation}
We can further denote the phonon specific heat as $c(\omega)$ and the mean free path as $l(\omega)$. The thermal conductivity\cite{berman1976} can thus be expressed as
\begin{equation}
\kappa = \frac{1}{3} \sum\limits_{\omega}^{}c(\omega){v}l(\omega) = \frac{1}{3}\sum\limits_{\omega}^{}c(\omega){v}^{2}\tau(\omega)
\end{equation}
So according to Eq. (12) and (13), the material with a higher defect density has the shorter relaxation time, and thus the lower thermal conductivity. For symmetric GBs in graphene, the density of 5$|$7 dislocation is determined by the orientation angle via the Frank formula, i.e. $\rho$ = 1/$d$ = 2$\mathrm{sin}$($\theta$/2)/$|\textbf{b}|$.\cite{song2013pseudo_model3} The GB with a larger orientation angle contains a higher 5$|$7 dislocation density. For the polycrystalline graphene with hexagonal grains as explored in this work, the density of defects increases as the grain size is reduced or the orientation angle increases. That is to say, the polycrystalline graphene with a smaller grain size or a larger orientation angle has the lower thermal conductivity, which agrees well with the results in Fig. 2. For other types of GBs, for example those with asymmetric GB or other defects, the scattering of heat flux is expected to reduce the thermal conductivity of polycrystalline graphene in a similar way as symmetric GBs we have discussed here.
\subsection{\label{sec:level2}The temperature-dependence of effective thermal conductivity}
First, we consider the temperature dependence of $\kappa$ of the pristine graphene. From the theory of phonon thermal transport we know that\cite{srivastava1991physics} the group velocity $v$ is determined by the phonon dispersion that does not change with temperature and the mean free path $l$ scales with the temperature $T$ as $1/T$,\cite{berman1976} and the specific heat $c$ should depends on the temperature $T$, namely $c(T)$, scales as $T^{3}$ at low temperature and approaches a constat when the temperature is close to the Debye temperature $T_\mathrm{D}$ (that is 1045 K for graphene).\cite{politano2011helium,pop2013MRS} Following Eq. (13) we can see that $\kappa \sim {T}^{-1}$ if the temperature dependence of $c$ is weak, which is very close to our simulation results $\kappa \sim {T}^{-0.96}$ and Cai's\cite{cai2010thermale2} experimental measurements $\kappa \sim {T}^{-1.03}$.

Then for the polycrystalline graphene that is treated as a composite material here,  we can assume that only two mechanisms contribute to the phonon relaxation or life time obtained from our MD simulations, those are (i) Umklapp scattering inside the grain (${\tau}_{\mathrm{U}}$) and (ii) the interfacial scattering at the GBs (${\tau}_{\mathrm{GB}}$). If we assume that these two processes are independent, we can invoke Matthiessen's rule and write the inverse lifetime of the phonon mode ($k$, $\omega$) a sum of contributions from each mechanism
\begin{equation}
\frac{1}{\tau(k,\omega)} = \frac{1}{{\tau}_{\mathrm{U}}(k,\omega)}+\frac{1}{{\tau}_{\mathrm{GB}}(k,\omega)}
\end{equation}
As temperature increases, the Umklapp scattering is enhanced as higher energy phonons are thermally populated\cite{kim2001thermal} and ${\tau}_{\mathrm{U}}$ is reduced.\cite{nika2009phonon} In contrast the GB scattering is weakened and ${\tau}_{\mathrm{GB}}$ increases.\cite{cao2012kapitza} Then as predicted by Eq. (13), $\kappa$ increases with the relaxation time. From the simulation results plotted in Fig. 5, we can see that $\kappa$ monotonically decreases with $T$, so we conclude that the Umklapp scattering plays a more important role than the GB scattering. Moreover, as the grain size decreases, the defect density increases and the GB scattering could make more contribution to the total relaxation time, which is, however, still lower than that from the Umklapp scattering. That is to say, the relaxation time $\tau$ of polycrystalline graphene with smaller grain size is reduced less than that of polycrystalline graphene with larger grain size as temperature increases, and the temperature dependence of $\kappa$ is weakened as the grain size decreases.
\section{\label{sec:level1}Conclusion}
In summary, we calculated the thermal conductivity $\kappa$ of polycrystalline graphene using equilibrium molecular dynamics simulations. It was found that the effective $\kappa$ increases with the grain size, and the results showed a linear dependence when the grain size is a few nanometers. On the other hand, the effective $\kappa$ can be reduced as the mismatch angle or dislocation density of the grain boundary increases. We also found that the effective $\kappa$ is reduced when temperature increase and the temperature dependence of $\kappa$ is much weakened when the grain size decreases. The underlying phonon scattering mechanisms were discussed through phonon scattering theories, with respect to the grain size, orientation, and temperature dependence. To predict the effective $\kappa$ with large grain size, we constructed an effective medium model, where polycrystalline graphene was considered as a two-phase composite. Based on the effective model we predicted $\kappa$ for grain size when grain size up to ten micrometers. The results showed that the thermal conductivity of polycrystalline graphene converges as the grain size increases to a few microns, that is the typical value in CVD-grown graphene.
\section{\label{sec:level1}Acknowledgment}
This work was supported by the National Natural Science Foundation of China through Grant 11222217, 11002079, Tsinghua University Initiative Scientific Research Program 2011Z02174, and the Tsinghua National Laboratory for Information Science and Technology of China.

\nocite{*}

\bibliography{reference}

\clearpage
\begin{figure}
\includegraphics[width=3.4in]{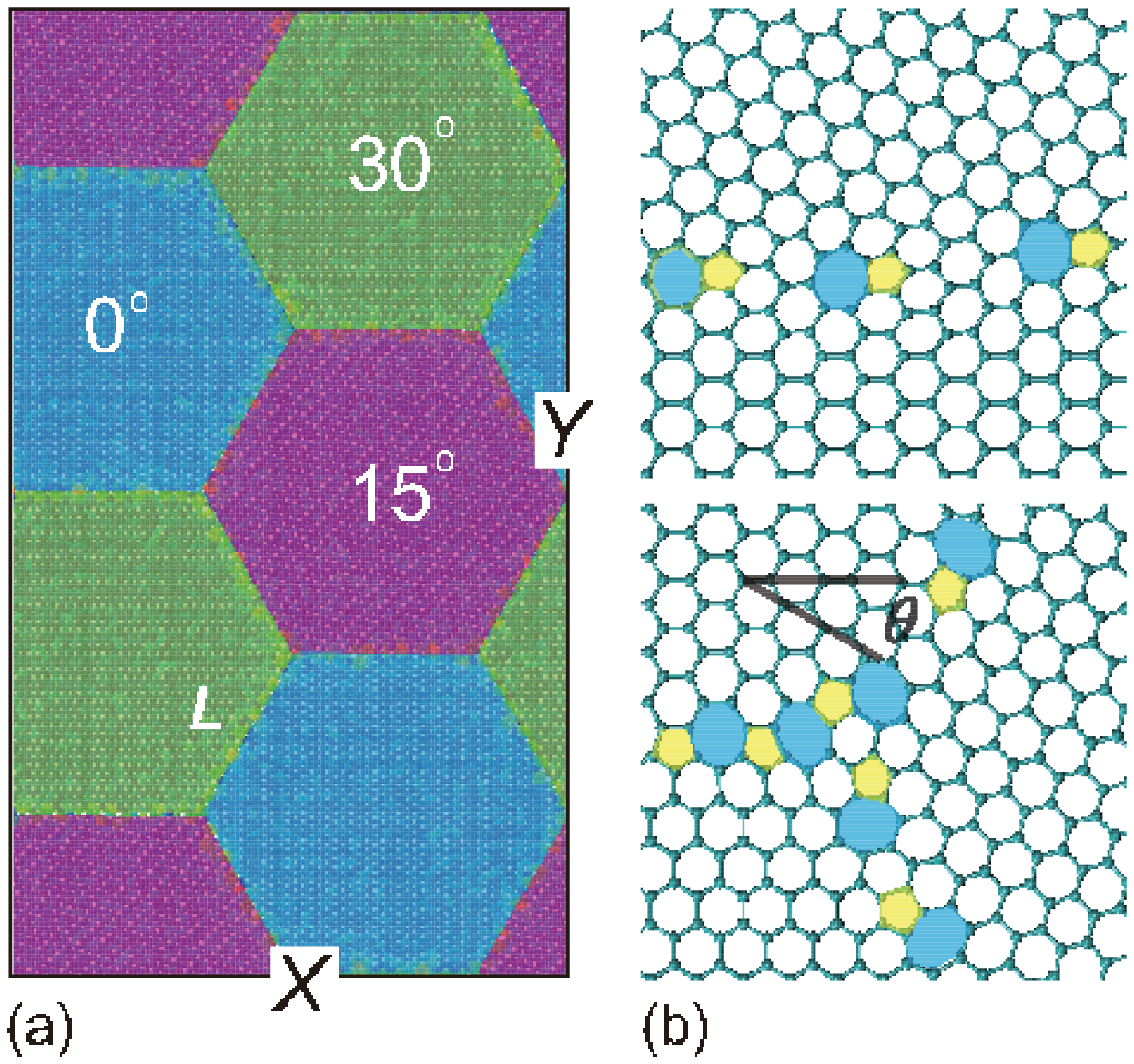}
\caption{\label{fig:epsart} (Color online) (a) The atomic structure of polycrystalline graphene where hexagonal grains of graphene with size \emph{L} and different orientations are patched into a supercell. Periodic boundary conditions are applied in both length and width directions. The lattice orientation of grains are denoted as $\theta$, that is (0\degree, 15\degree, 30\degree) in this illustration. (b) The grain boundaries are built up by pentagons, hexagons and heptagons, and the typical atomic structure and intersection are plotted.}
\end{figure}

\clearpage
\begin{figure}
\includegraphics[width=3.4in]{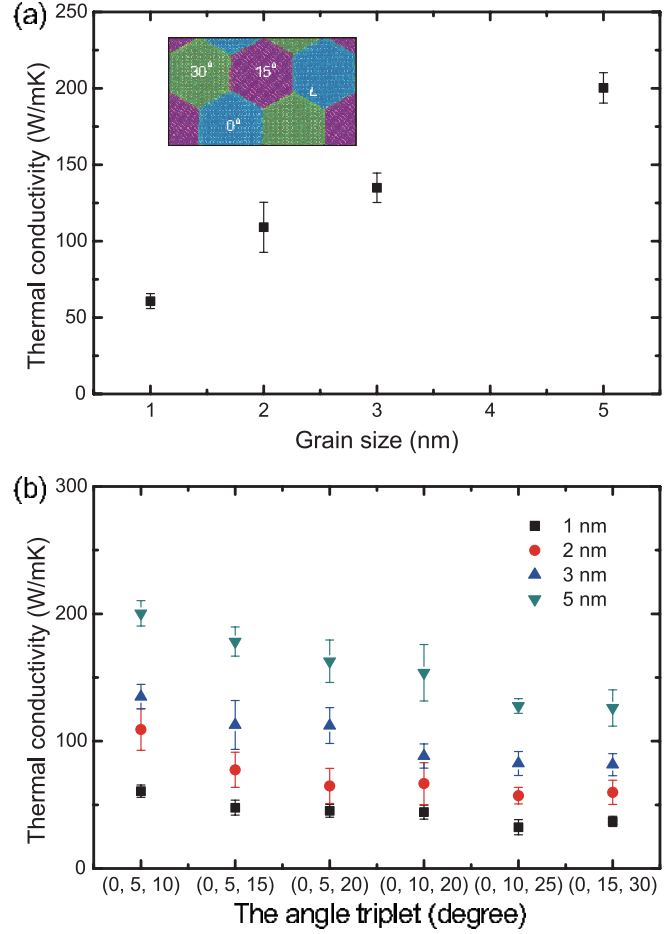}
\caption{\label{fig:epsart} (Color online) (a) Thermal conductivity of polycrystalline graphene as a function of the grain size \emph{L} for the mismatch angle triplet (0\degree, 5\degree, 10\degree). (b) Thermal conductivity of polycrystalline graphene as a function of the relative orientation angle ${\theta}_{r}$ for different grain size, which are 5\degree, 10\degree, 13.33\degree, 13.33\degree, 16.67\degree\ and 20\degree\ from left to right, respectively.}
\end{figure}

\clearpage
\begin{figure}
\includegraphics[width=3.4in]{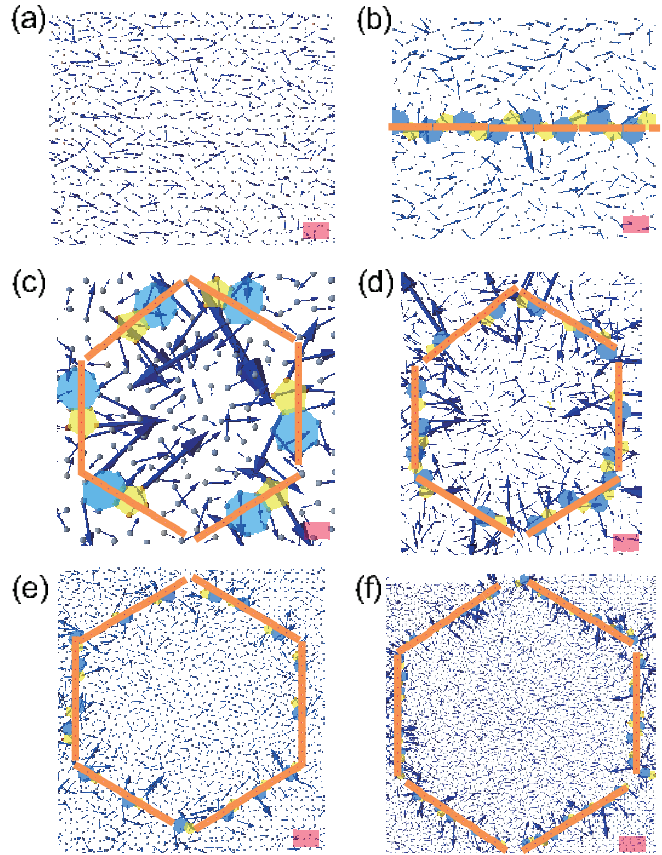}
\caption{\label{fig:epsart} (Color online) Spatial distribution of the heat flux in pristine and polycrystalline graphene for the specified angle triplet (0\degree, 5\degree, 10\degree). (a) Pristine graphene. (b) Polycrystalline graphene with single grain boundary across the periodic boundary, and the heat flux is parallel to the grain boundary. (c-f) Polycrystalline graphene with grain size of 1, 2, 3 and 5 nm, respectively. The length of scale bar is 0.49, 0.15, 0.14, 0.20, 0.31, 0.43 nm, respectively.}
\end{figure}

\clearpage
\begin{figure}
\includegraphics[width=3.4in]{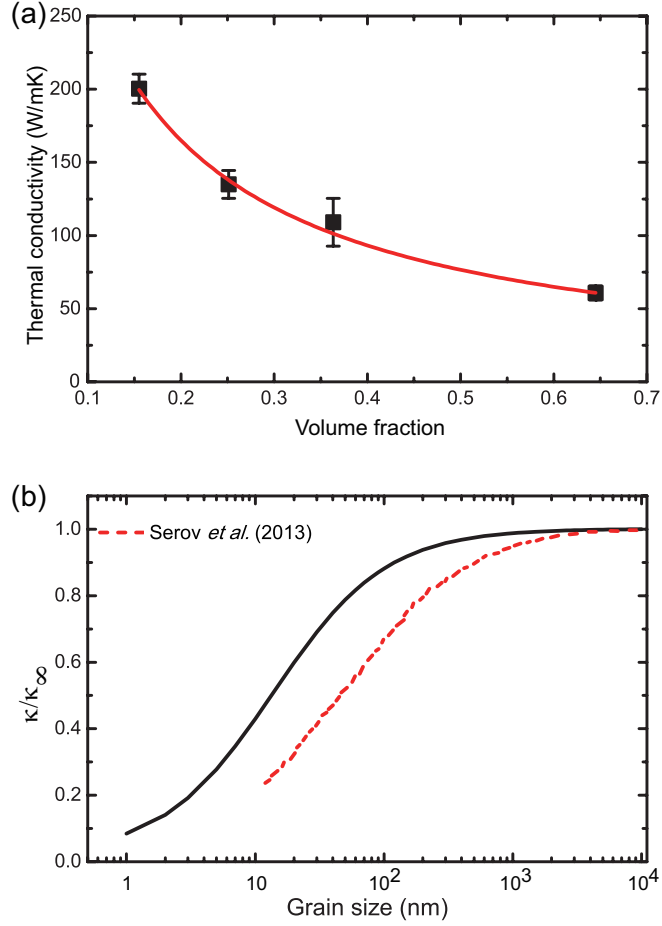}
\caption{\label{fig:epsart} (Color online) (a) Thermal conductivity of polycrystalline graphene as a function of the volume fraction of grain boundary phase ${f}_{\mathrm{GB}}$ for the specified angle triplet (0\degree, 5\degree, 10\degree). We fitted the results via the effective medium model and the result shows $\kappa = 719.97/(1+16.81{f}_{\mathrm{GB}})\  \mathrm{W/mK}$. (b) The ratio ${\kappa}$/${\kappa}_{\infty}$ as a function of the grain size for the specified angle triplet (0\degree, 5\degree, 10\degree). The black line is our prediction and the red dashed line is data taken from Serov \emph{et al.}\cite{grosse2011pop}}
\end{figure}

\clearpage
\begin{figure}
\includegraphics[width=3.4in]{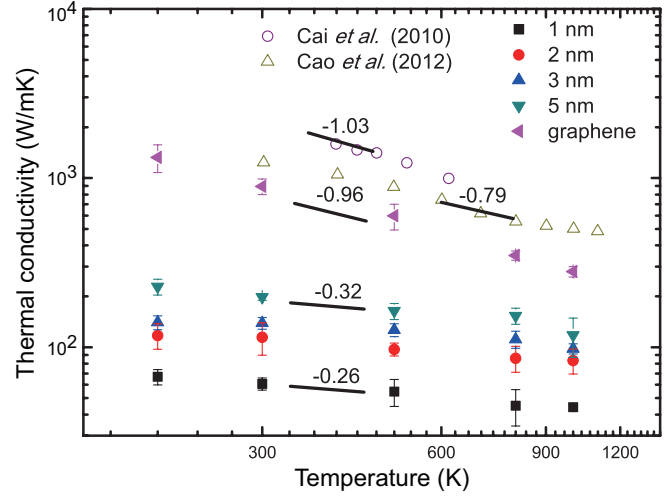}
\caption{\label{fig:epsart} (Color online) Thermal conductivities of pristine and polycrystalline graphene as a function of temperature and grain size for the specified angle triplet (0\degree, 5\degree, 10\degree). Experimental and simulation results for the temperature dependence found in the literature\cite{cai2010thermale2,cao2012kapitza} are also plotted here for comparison.}
\end{figure}

\clearpage
\begin{figure}
\includegraphics[width=3.4in]{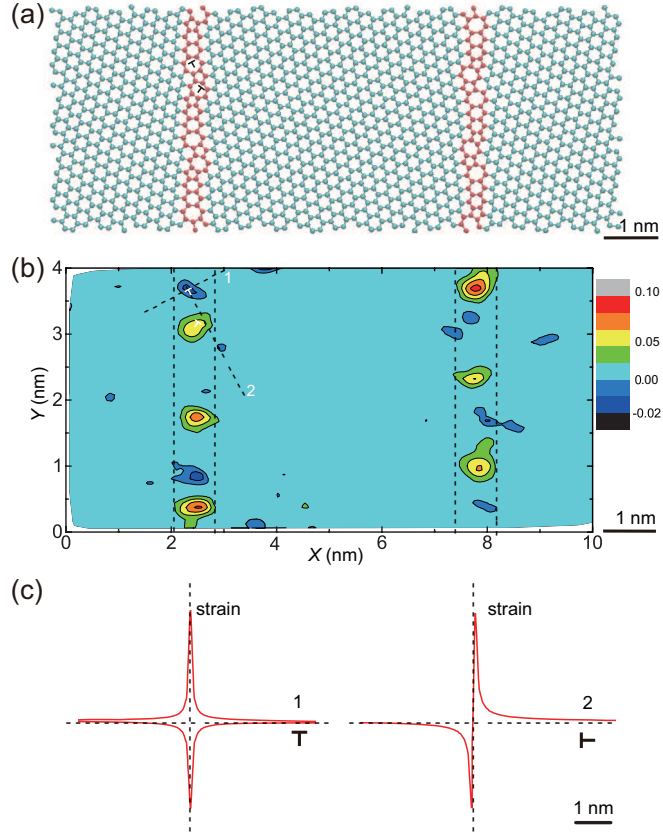}
\caption{\label{fig:epsart} (Color online) (a) The atomic structure of graphene with parallel grain boundaries. The mismatch angle is 15.4\degree. (b) The spatial distribution of strain field in the graphene with an armchair GB, calculated using the data from our MD simulations. (c) The shear strain field induced by a dislocation calculated using Eq. (8), plotted along lines 1 and 2 as denoted in panel (b).}
\end{figure}

\end{document}